# Prediction of COVID-19 using chest X-ray images

Narayana Darapaneni[1], Suma Maram[2], Harpreet Singh[3], Syed Subhani[4], Mandeep Kour[5], Sathish Nagam[6], and Anwesh Reddy Paduri[7]

*[1] Northwestern University/Great Learning, Evanston, US*

*[2-7]Great Learning, Bangalore, India*

*anwesh@greatlearning.in*

**Abstract:** COVID-19, also known as Novel Coronavirus Disease, is a highly contagious disease that first surfaced in China in late 2019. SARS-CoV-2 is a coronavirus that belongs to the vast family of coronaviruses that causes this disease. The sickness originally appeared in Wuhan, China in December 2019 and quickly spread to over 213 nations, becoming a global pandemic. Fever, dry cough, and tiredness are the most typical COVID-19 symptoms. Aches, pains, and difficulty breathing are some of the other symptoms that patients may face. The majority of these symptoms are indicators of respiratory infections and lung abnormalities, which radiologists can identify. Chest x-rays of COVID-19 patients seem similar, with patchy and hazy lungs rather than clear and healthy lungs. On x-rays, however, pneumonia and other chronic lung disorders can resemble COVID-19. Trained radiologists must be able to distinguish between COVID-19 and an illness that is less contagious. Our AI algorithm seeks to give doctors a quantitative estimate of the risk of deterioration. So that patients at high risk of deterioration can be triaged and treated efficiently. The method could be particularly useful in pandemic hotspots when screening upon admission is important for allocating limited resources like hospital beds.

## 1. Introduction

At the end of 2019, humankind was faced with an epidemic—severe acute respiratory syndrome coronavirus (SARS CoV-2)–related pneumonia, referred to as coronavirus disease 2019 (COVID-19)—that people did not expect to encounter in the current era of technology. While the COVID-19 outbreak started in Wuhan, China, the significant spread of the epidemic around the world has meant that the amount of equipment available to doctors fighting the disease is insufficient. Considering the time required for diagnosis and the financial costs of the laboratory kits used for diagnosis, artificial intelligence (AI) and deep learning research and applications have been initiated to support doctors who aim to treat patients and fight the illness. The financial costs of the laboratory kits used for diagnosis, especially for developing and underdeveloped countries, are a significant issue when fighting the illness. Using X-ray images for the automated detection of COVID-19 might be helpful for countries and hospitals that are unable to purchase a laboratory kit for tests or that do not have a CT scanner. This is significant because, currently, no effective treatment option has been found, and therefore effective diagnosis is critical

## 2. Literature survey

The foremost aim of this automated system using machine learning is to analyze the characteristics of disease and provide some valuable predictions. Thus, the main steps are image pre-processing, segmentation of the disease-related regions of interest, computing effective features and building feature-based machine learning models for detection and classification. For example, a classification of COVID-19 & non COVID-19 cases using KNN model provides an accuracy of 96.4% [17]this number is not matching. Many DL models have been included in the literature to classify and detect COVID-19 cases here. The proposed method uses deep learning to predict covid's from chest X-ray images. This scheme divides images into two categories: COVID-19-infected and non-COVID-19-infected. Chest x-ray (i.e., radiography) and chest CT are a more effective imaging technique for diagnosing lung related problems. Still, a substantial chest x-ray is a lower cost process in comparison to chest CT. COVID-19 X-ray images have indicated opacity-related findings [1]. Bilateral and unilateral ground-glass opacity was found in one study's patients [2]. Consolidation and ground-glass opacities were found in 50-60% of COVID-19 cases in pediatric patients [3]. This key characteristic may be useful in developing a deep learning model to facilitate in screening of large volumes of radiograph images for COVID-19 suspect cases.

Deep learning is the most successful machine learning technique, providing useful analysis to study a large number of chest x-ray images, which can have a significant impact on Covid-19 screening

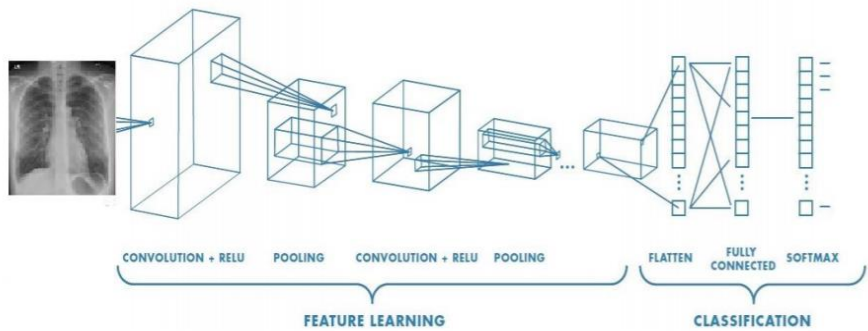

Fig.1 System overview

In this work, we have taken the PA view of chest x-ray scans for covid-19 affected patients as well as healthy patients. After cleaning up the images and applying data augmentation, we will be experimenting with deep learning-based CNN models and comparing their performance. Deep learning has the potential to revolutionize the automation of lung radiography interpretation. More than 40,000 research articles have been published related to the use of deep learning in this topic including the establishment of referent data set [4], organ segmentation [5], artefact removal [6], multilabel classification [7], data augmentation [8], and grading of disease severity [9]. The key component in deep learning research is the availability of training and testing dataset, whether it is accessible to allow reproducibility and comparability of the research.

Transfer learning is a technique widely used in deep learning that allows previously trained models to be reused in a particular application [7]. As seen in the ImageNet database [10], proven pre-trained deep neural networks have been trained on at least a million images to recognize thousands of objects. The image set consists of typical and atypical objects, for example, pencil, animals, buildings, fabrics, and geological formation. One method of transfer learning is to freeze all layers except the last three layers—fully connected, SoftMax, and classification layers. The last three layers are then trained to recognize new categories. Pretrained models have shown promising results, in some instances, comparable with experienced radiologists [11]. Fig.2 shows the architecture of the transfer learning model.

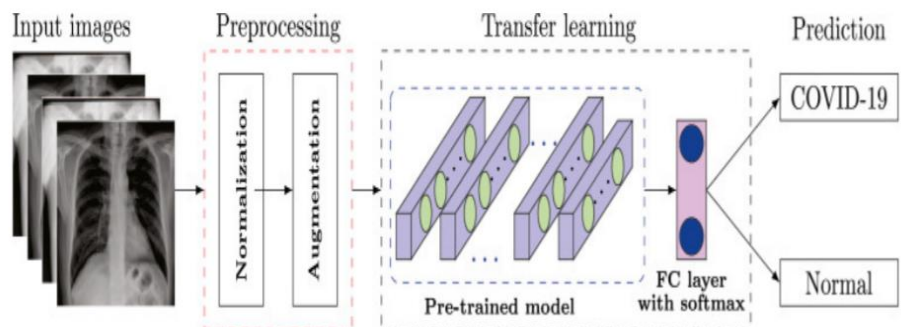

Fig.2 Transfer learning

For deep learning to be effective, data quality is critical. The idiom "garbage in, garbage out" applies as well to general deep learning applications as it does to lung radiography deep learning. According to previous studies, radiologist interpretive errors are caused by both internal and external factors [12]. Scan, recognition, judgement, and cognitive errors are examples of the former, while exhaustion, workload, and distraction are examples of the latter. Underperforming models will result from inaccurate labels used to train deep learning architecture.

In recent research [11], radiologist-adjudicated labels for the lung X-ray 14 data set [4] were created. These labels are unusual in that they needed adjudication by a group of licensed radiologists with more than 3 years of general radiology experience. Four labels were introduced, namely, pneumothorax, nodule/mass, airspace opacity, and fracture. With the recent discovery of opacity as a significant feature in COVID-19 patients, this study aims to build a deep learning model for COVID-19 case prediction based on an established pretrained model that was then retrained using adjudicated data to recognize images with airspace opacity, a COVID-19 abnormality.

## 3. Step-by-step walk through of the solution

Steps that are followed are as follows:

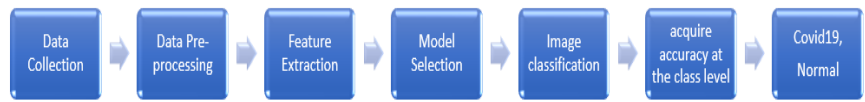

- ❖ The data for this study came from the Kaggle repository, which includes Chest X-Ray images of Covid-19 affected, normal, and pneumonia patients. Our model was developed and evaluated using a Dataset of 21,025 chest X-ray exams. Examples of chest X-ray images acquired from various patients are shown in Fig 1.a and 1.b. For model construction and hyperparameter tuning, a dataset of 2000 chest X-ray pictures was used. There was no patient crossover between the training and test sets.
- ❖ Resize the input images to match the scale of the input layer of the pre-trained network's input layer. Sample Covid x-ray images are shown in fig.,3(a) and fig.,3(b).
- ❖ To get reliable results from a deep learning method, you'll need a significant amount of data. However, it is likely that every problem lacks sufficient data. Collecting data, particularly in medical-related issues, can be time consuming and expensive. Augmentation can be used to solve these types of problems. Augmentation can overcome the problem of overfitting and enhance the accuracy of the proposed model.
- ❖ Divide the data into training and test sets; 80 percent of the photos will be utilized for training, while the remaining 20% will be used to test the network.
- ❖ To identify the probability of COVID-19 and normal class, modify the network architecture by swapping the final layers of the pre-trained network as: "average pooling", "fully-connected layer", "softmax" with a "classification output".
- ❖ Train the Network.

❖ Test the classifiers on the testing dataset. On a collected dataset of chest X-ray pictures, we used three models (VGG-16, Resnet-50, InceptionV3)

❖ Plot the accuracy and loss during the training and analyze the performance of the model.

❖ Design UI to detect covid based on radiographic image provided.

### Positive COVID-19 Chest X-ray

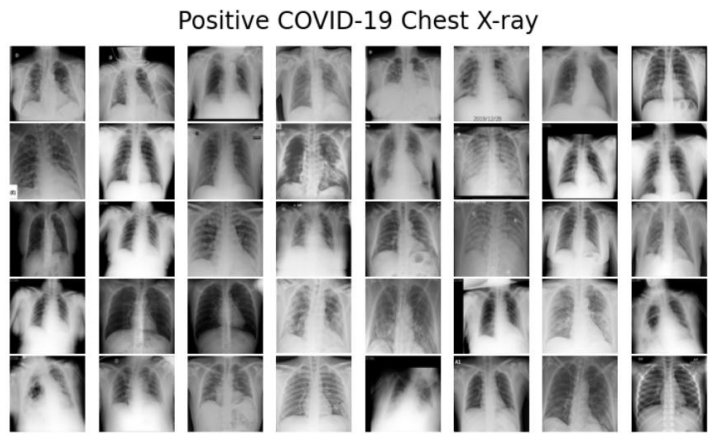

Fig.3.a.PositiveCovid-19X-ray

### Negative COVID-19 Chest X-ray

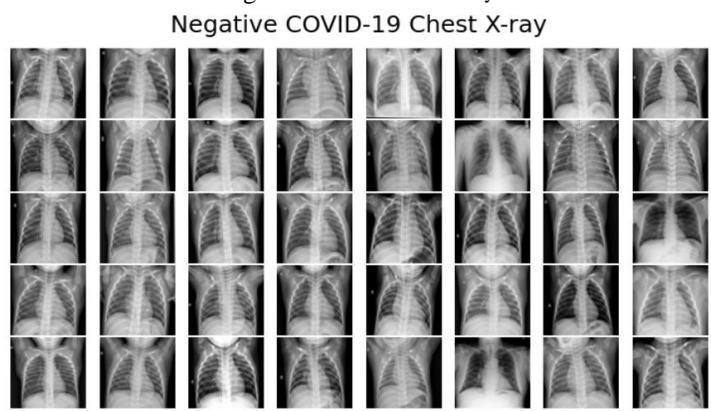

Fig.3.b Normal (Covid19 Negative) X-ray

## 4. Model evaluation

### 4.1 VGG16

Karen Simonyan and Andrew Zisserman developed the first VGG models, which were published in the paper Very Deep Convolutional Networks for Large-Scale Image Recognition in 2015. VGG16 has 16 weighted layers, while VGG99 has 19 weighted layers.

The VGG architecture is simple and similar to that of the original convolutional networks. VGG's major goal was to make the network deeper by stacking additional convolutional layers on top of each other. This was made possible by limiting the convolutional windows to only 3x3 pixels in size.

The classifier is trained using ImageNet's 1000 categories. However, because our objective is to categorize Covid and Normal X-ray images, we only have two categories. By changing the include top parameter to False, it's simple to import only the convolutional component of VGG16.

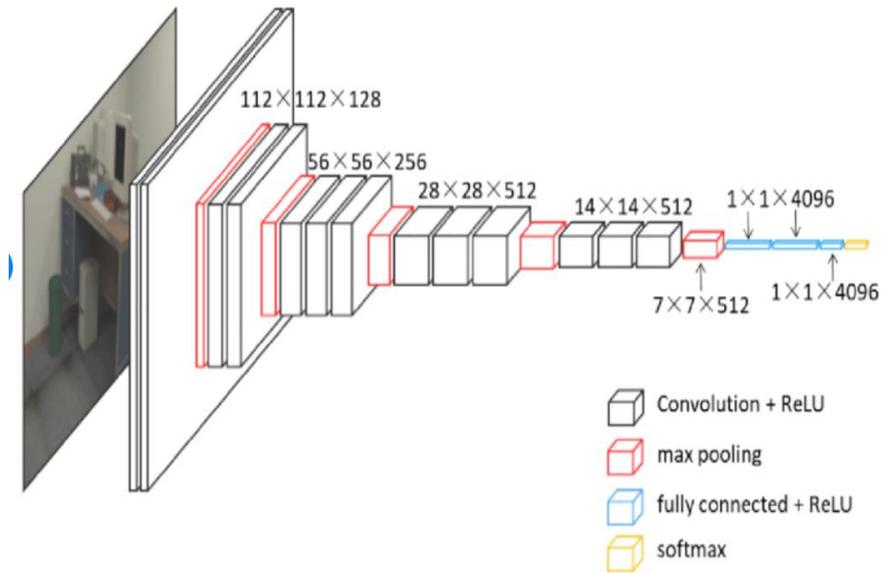

Architecture of VGG16.

The ROC curve, Classification report, confusion matrix, Model Accuracy and Model Loss plots are given below.

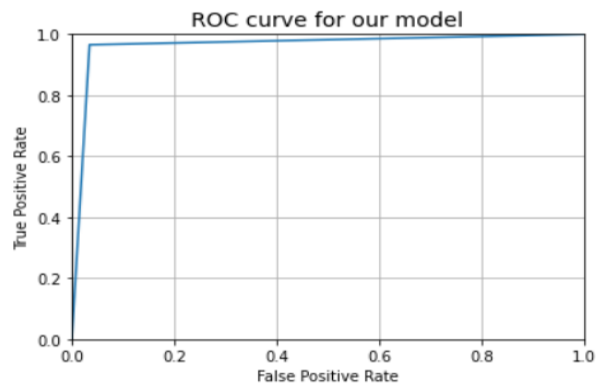

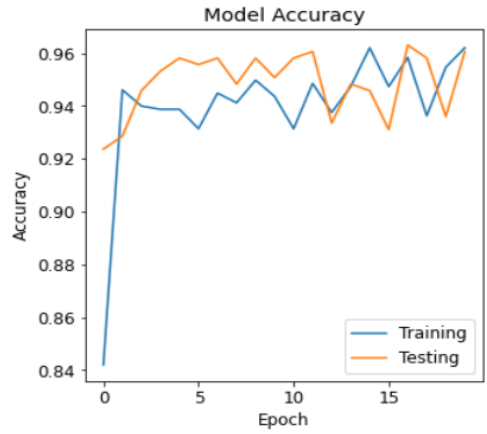

**Fig.4.** Accuracy vs Epocs(VGG16)

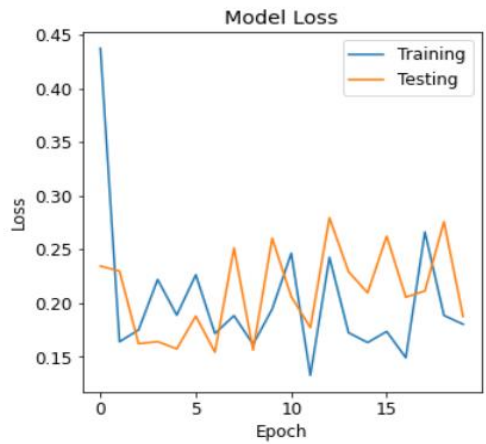

Fig.5. Loss vs Epocs(VGG16)

|  | precision | recall | f1-score | support |
|---|---|---|---|---|
| 0 | 0.97 | 0.97 | 0.97 | 203 |
| 1 | 0.97 | 0.97 | 0.97 | 203 |
| accuracy |  |  | 0.97 | 406 |
| macro avg | 0.97 | 0.97 | 0.97 | 406 |
| weighted avg | 0.97 | 0.97 | 0.97 | 406 |

**Fig.6.** Classification Report for VGG16

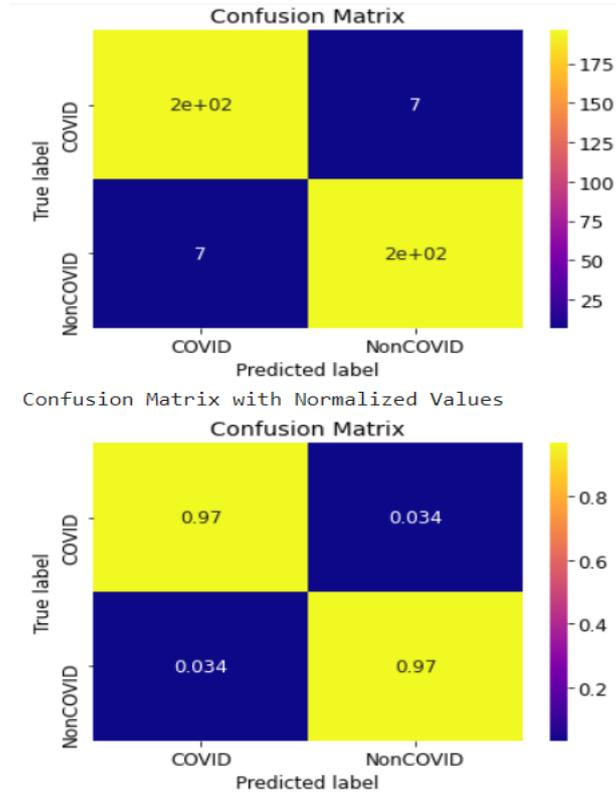

Confusion Matrix with Normalized Values

## 4.2 ResNet50

ResNet-50 is a 50-layer deep convolutional neural network. You can use the ImageNet dataset to load a pre-trained version of the network that has been trained on over a million photos. The network can classify photos into 1000 different object categories, including keyboards, mice, pencils, and a variety of animals.

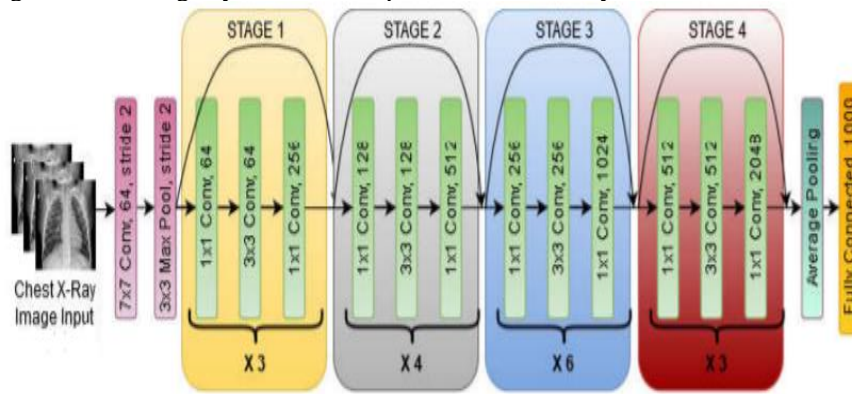

**Fig.7.** Architecture of ResNET50

The ROC curve, Classification report, confusion matrix, Model Accuracy and Model Loss plots are given below.

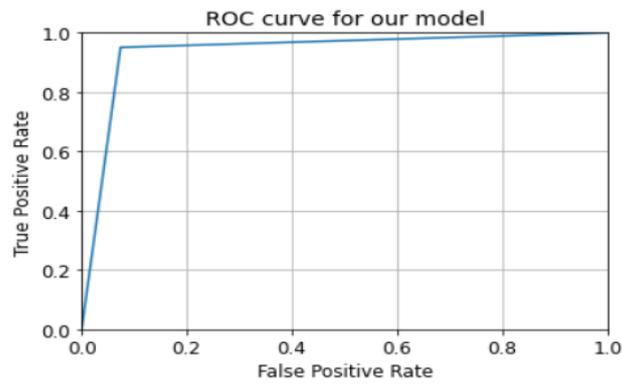

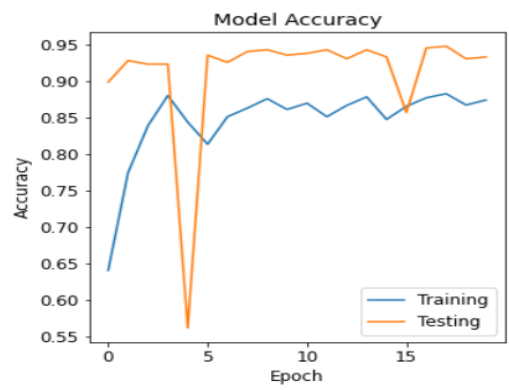

Fig.8. Accuracy vs Epocs(ResNet50)

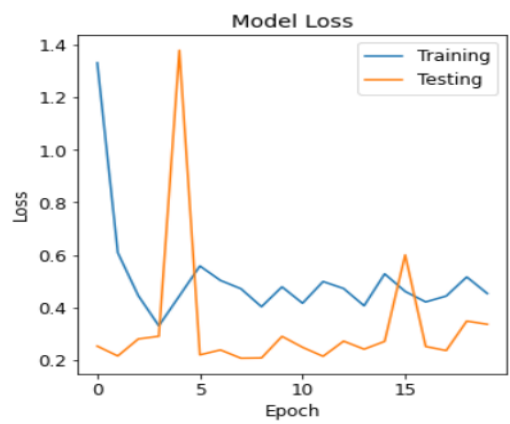

Fig.9. Loss vs Epocs(ResNet50)

|              | precision | recall | f1-score | support |
|--------------|-----------|--------|----------|---------|
| 0            | 0.95      | 0.93   | 0.94     | 203     |
| 1            | 0.93      | 0.95   | 0.94     | 203     |
| accuracy     |           |        | 0.94     | 406     |
| macro avg    | 0.94      | 0.94   | 0.94     | 406     |
| weighted avg | 0.94      | 0.94   | 0.94     | 406     |

Fig.10. Classification Report for ResNet50

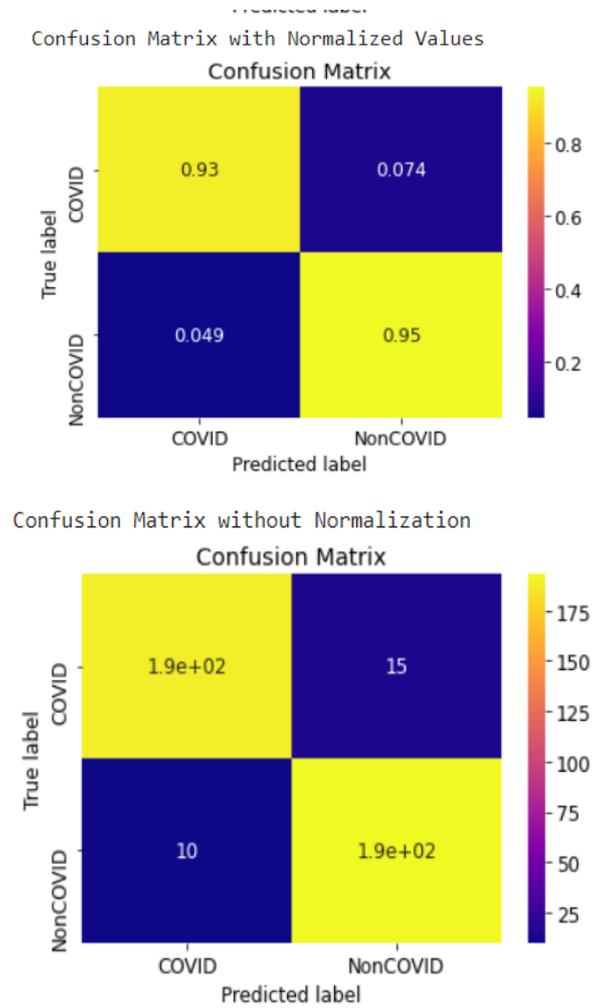

### 4.3 InceptionV3

Inception-V3 is a 48-layer deep convolutional neural network. You can use the ImageNet database to load a pre-trained version of the network that has been trained on over a million photos.

The network can classify photos into 1000 different object categories, including keyboards, mice, pencils, and a variety of animals. As a result, the network has learned extensive feature representations for a variety of image types.

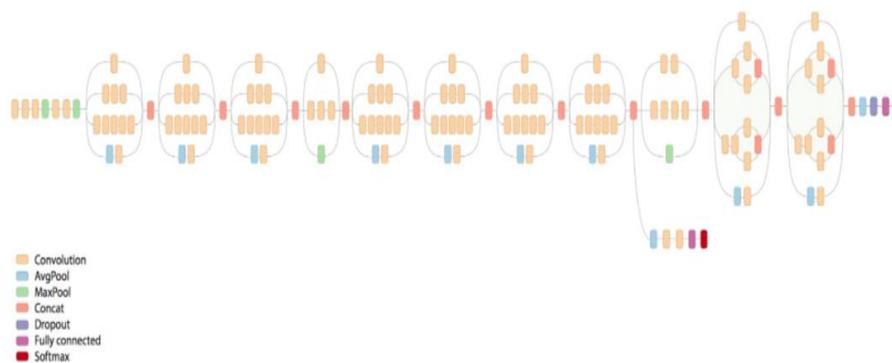

Fig.11. Architecture of InceptionV3

The ROC curve, Classification report, confusion matrix, Model Accuracy and Model Loss plots are given below.

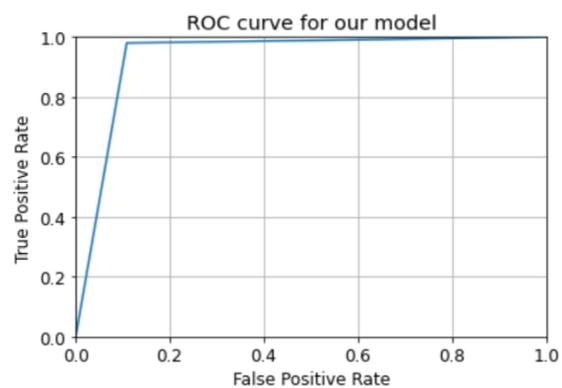

|              | precision | recall | f1-score | support |
|--------------|-----------|--------|----------|---------|
| 0            | 0.98      | 0.89   | 0.93     | 203     |
| 1            | 0.90      | 0.98   | 0.94     | 203     |
| accuracy     |           |        | 0.94     | 406     |
| macro avg    | 0.94      | 0.94   | 0.94     | 406     |
| weighted avg | 0.94      | 0.94   | 0.94     | 406     |

Confusion Matrix with Normalized Values

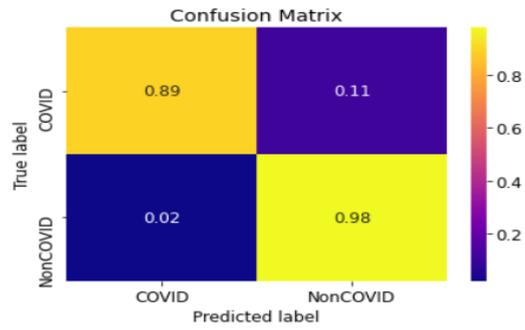

Confusion Matrix without Normalization

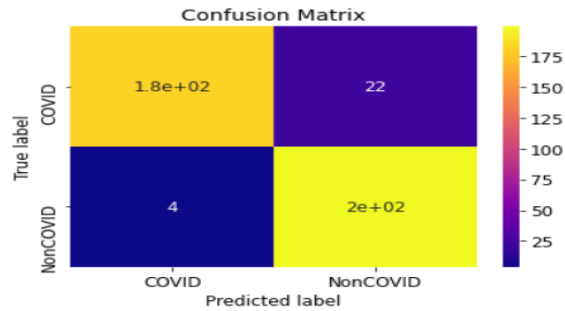

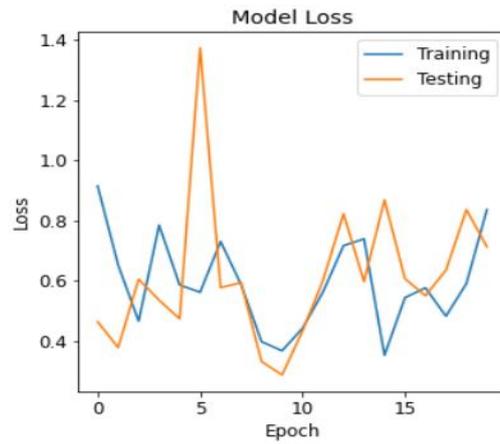

## 5. Comparison to benchmark :

| Model | No. of parameters | Testing Accuracy |
| --- | --- | --- |
| VGG-16 | 20,074,562 | 97 |
| ResNet50 | 23,788,418 | 94 |
| Inception V3 | 23,788,418 | 94 |

As we can observe from our models' accuracies, The maximum accuracy was attained with VGG-16, followed by RestNet50 and inceptionV3.

## 6. Visualization(s)

### 6.1 VGG16

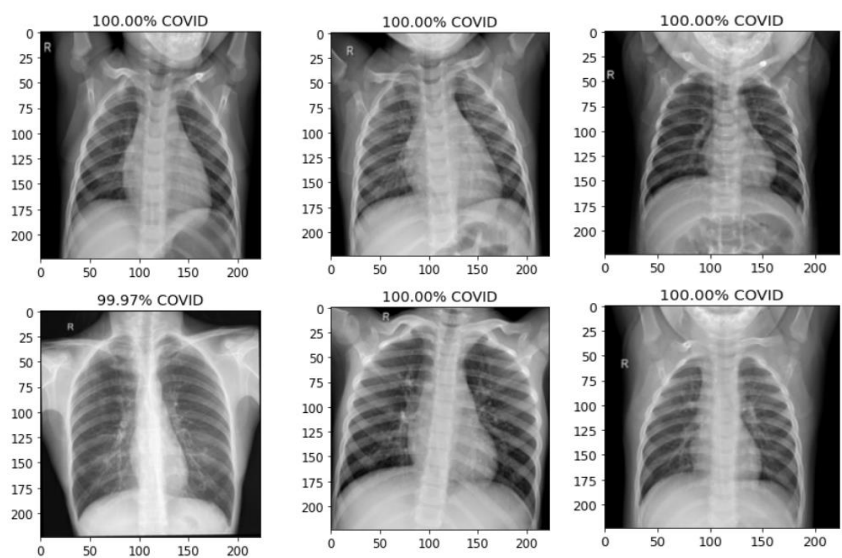

## 6.2 ResNet50

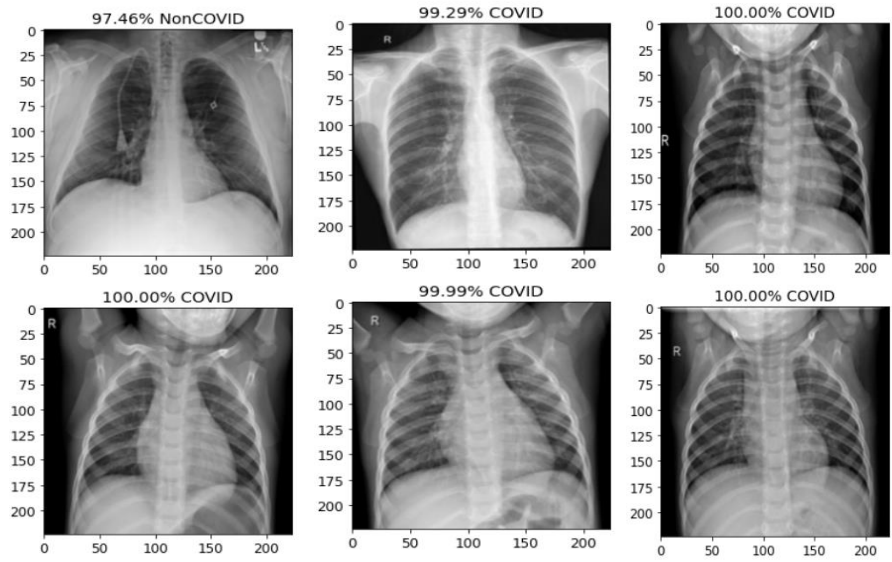

## 6.3 InceptionV3

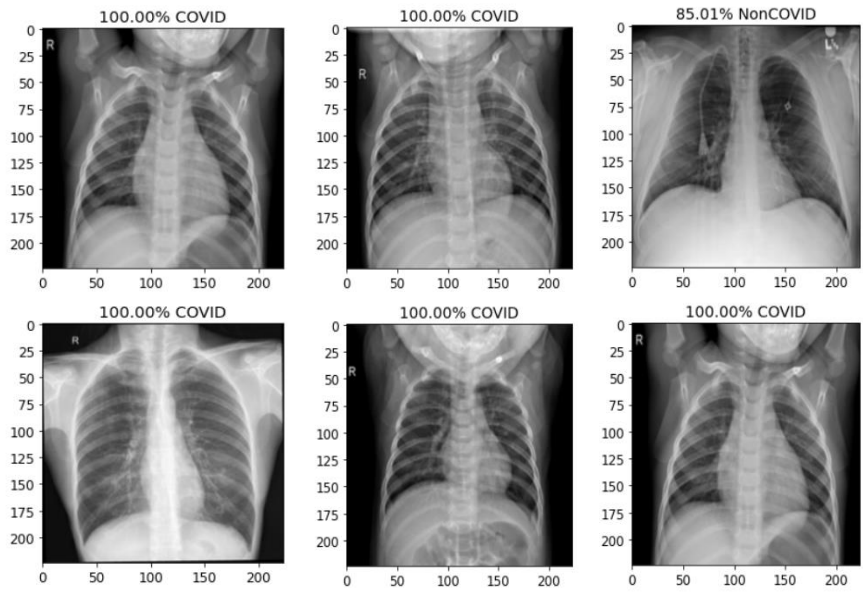

## 7. GUI

Compared to RestNet and Inception models built, VGG model accuracy, precision and recall scores found to be providing best results. A Flask based UI is built that uses radiographic image as input and provides the prediction using VGG Model built. Prediction output will be a binary classification to detect covid/normal based on the image selected.

It contains Flask APIs that receives radiographic image details through GUI or API calls, computes the predicted value based on vgg and returns the prediction result. It has the HTML template and CSS styling to allow users to browse image details from client location. It sends the image data via REST API(/detect) using POST method and displays the result whether covid positive/normal.

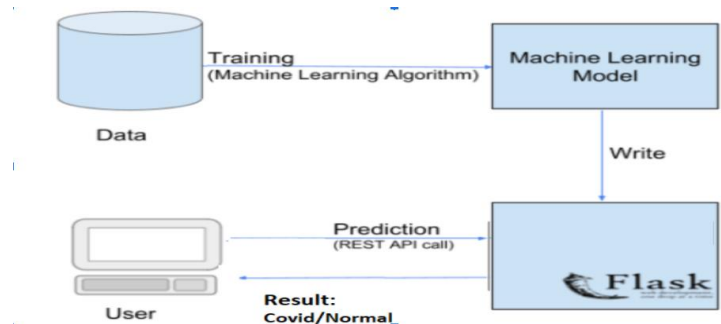

Fig.12. Pipeline for deployment of model including UI

Initial Look. Enter patient info(Optional). Select image using choose file option

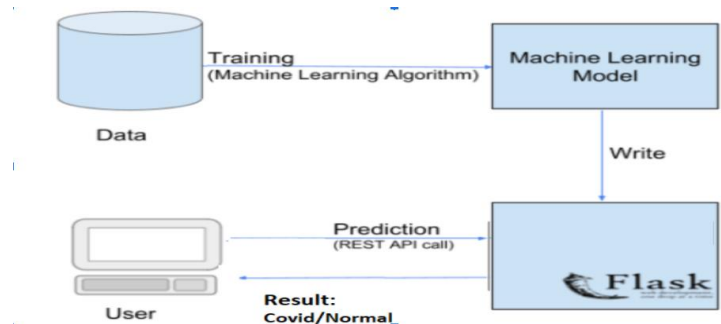

Fig.13. Select image from covid test images. Result is detected as COVID

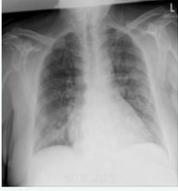

Select a normal image. Result is detected as normal.

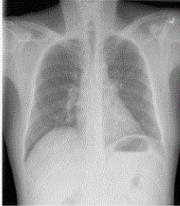

## 8. API Sample Usage

### 8.1 Covid Image:

### 8.2 Normal Image:

## 9. Implications

In this study we proposed a deep learning-based algorithm to detect and classify COVID-19 instances from X-ray images. Our model has an end-to-end structure that eliminates the need for manual feature extraction. Our designed method is capable of performing with an accuracy rate of 97%. This technique can be used to solve a scarcity of radiologists in rural areas of countries affected by COVID-19.

We plan to add more images to our model in the future to validate it. This generated model can be stored in the cloud to give fast diagnosis and aid in the rehabilitation of affected patients. This should considerably minimize clinician effort.

## 10. Limitations

To get reliable results from a deep learning method, you'll need a significant amount of data, However, it is likely that every problem lacks sufficient data. Collecting data, particularly in medical-related issues, can be time consuming and expensive. In the future, we intend to make our model more robust and accurate by using more such images.

## 11. Closing Reflections

We demonstrated the ability of AI to aid in the successful diagnosis of COVID-19 utilizing X-ray images using a commercial platform that is freely available. While this technique has a wide range of applications in radiology, we've concentrated on its potential impact on future global health issues like COVID-19. The findings have implications for patient screening and triage, early diagnosis, illness progression tracking, and identifying patients who are at higher risk of morbidity and mortality. Our research provides a glimpse into how artificial intelligence (AI) will likely transform medical practice in the future.